\pdfoutput=1
\documentclass[english]{article}
\usepackage{geometry}
\usepackage{babel}
\usepackage{xcolor}
\usepackage{graphicx}
\usepackage{amsmath,amsfonts}
\usepackage{xcolor}
\usepackage{mathcomp}
\usepackage{textcomp}
\usepackage{amssymb}
\usepackage{graphicx,indentfirst,subfigure,epsfig}
 \usepackage{varioref}
 \usepackage{wrapfig}
 \usepackage{subfigure}
 \usepackage{subfigmat}
 \usepackage{amsthm}
\usepackage{bm}

\graphicspath{{./}{./figs/}}

  \def\clap#1{\hbox to 0pt{\hss#1\hss}}

\providecommand{\mat}[1]{\bm{#1}}%
\renewcommand{\vec}[1]{\mathbf{#1}}

\providecommand{\mA}{\ensuremath{\mat{A}}}

\providecommand{\mC}{\ensuremath{\mat{C}}}

\providecommand{\mI}{\ensuremath{\mat{I}}}

\providecommand{\mU}{\ensuremath{\mat{U}}}

\providecommand{\mW}{\ensuremath{\mat{W}}}

\providecommand{\vc}{\ensuremath{\vec{c}}}

\providecommand{\vh}{\ensuremath{\vec{h}}}

\providecommand{\vu}{\ensuremath{\vec{u}}}

\providecommand{\vx}{\ensuremath{\vec{x}}}

\usepackage{hyperref}
\hypersetup{colorlinks, citecolor=red, linkcolor=blue, urlcolor=red, breaklinks=true}
\usepackage{scrextend}
\usepackage[T1]{fontenc}
\usepackage[utf8]{inputenc}
\usepackage{tgtermes}
\usepackage{nomencl}
\usepackage{algorithm}
\usepackage{algorithmic}
\usepackage{booktabs}
\usepackage{longtable}
\usepackage{lineno}
\makenomenclature

\begin{document}
\large
\title{\textbf{AeroVR: Immersive Visualization System for Aerospace Design}}
\author{S\l awomir Konrad Tadeja\footnote{Address all correspondence to \texttt{skt40@cam.ac.uk}}$\;^{\; \star}$, Pranay Seshadri$^{\star}$, Per Ola Kristensson$^{\star}$\\ \vspace{0.3 cm} \\ $^{\star}$Department of Engineering, University of Cambridge, Cambridge, U. K.}
\date{}
\maketitle{}
\begin{abstract}
One of today's most propitious immersive technologies is \textit{virtual reality} (VR). This term is colloquially associated with headsets that transport users to a bespoke, built-for-purpose immersive 3D virtual environment. It has given rise to the field of immersive analytics---a new field of research that aims to use immersive technologies for enhancing and empowering data analytics. However, in developing such a new set of tools, one has to ask whether the move from standard hardware setup to a fully immersive 3D environment is justified---both in terms of efficiency and development costs. To this end, in this paper, we present the \textit{AeroVR}--an immersive aerospace design environment with the objective of aiding the component aerodynamic design process by interactively visualizing performance and geometry. We decompose the design of such an environment into function structures, identify the primary and secondary tasks, present an implementation of the system, and verify the interface in terms of usability and expressiveness. We deploy AeroVR on a prototypical design study of a compressor blade for an engine.
\end{abstract}

\section{INTRODUCTION}
Virtual reality (VR) is rapidly being hailed as the new paradigm for interactive visualization of data. Its ability to fuse visual, audio, and haptic sensory feedback in a computer-generated simulation environment is deemed to have tremendous potential. While the phrase \emph{virtual reality} has been used for decades, in the context of computer aided visualization, today it is synonymous with head-mounted displays \cite{sutherland_head-mounted_1968} (HMDs) or headsets \cite{oculus, nokuo2014head}. Although still in a nascent stage, HMDs have demonstrated their usefulness in the computer gaming, education, fashion and real-estate industries, with countless more application areas currently being pursued, including information visualization in aerospace \cite{garcia-hernandez_perspectives_2016}. One potentially promising application is aerospace design---a complex, multi-disciplinary, multi-objective and multi-dimensional problem---where technologies that offer faster design cycle times, with potentially greater efficiency gains, can be real game-changers. However, as the aerospace community usually works on state-of-the-art computational tools and sophisticated computer-aided design packages, there are tremendous hurdles in getting the community to embrace VR. Furthermore, at this stage, it is not precisely clear what the benefits are in migrating to a VR-based design framework. Thus, what is required by the aerospace design community is an initial sketch of an immersive VR aerospace design environment---the AeroVR, a computer-generated environment that leverages the full visual, audio, and haptic sensory frameworks afforded by VR technology.

Our focus in this paper is to explore how aerospace design workflows can benefit from VR. To aid our effort, we will be using ideas from parameter-space dimension reduction \cite{constantine2015active, cook2009regression, seshadri2019supporting}. This topic has recently received considerable attention from both the applied mathematics and computational engineering communities, where the aim has been to reduce the cost of expensive computer parameter studies---that is, optimization, uncertainty quantification, and more generally \emph{design of experiments}. In \ref{sec:dim}, we present some of the key theoretical ideas that underpin dimension reduction. This is followed in section \ref{sec:aero} with a presentation of the VR aerospace design environment including its function and system structures, tasks analysis, and interaction features. The next section \ref{sec:eval} describes a verification of the interface with respect to usability and expressiveness. Section \ref{sec:future} summarizes the contributions and outlines future work.

\section{PARAMETER-SPACE DIMENSION REDUCTION} \label{sec:dim}
Consider a function $f(\vx)$ where $f: \mathbb{R}^{d} \rightarrow \mathbb{R}$. Here $f$ represents our chosen quantity of interest (qoi); the desired output of a computational model. This qoi can be the lift coefficient of a wing or indeed the efficiency of a turbomachinery blade. Let $\vx \in \mathbb{R}^{d}$ be a vector of design parameters. Now when $d \leq 2$, visualizing the design space of $f$ is trivial, one needs to simply run a design of experiment and view the results as a scatter plot. However, when $d \geq 3$ visualizing the design space becomes difficult. One way forward is to approximate $f$, with 
\begin{equation}
f(\vx) \approx g(\mU^T \vx),
\label{ridge_def}
\end{equation}
where $g: \mathbb{R}^{m} \rightarrow \mathbb{R}$ and $\mU \in \mathbb{R}^{d \times m}$, with $m \leq d$. We call the subspace associated with the span of $\mU$ its \emph{ridge subspace} and $g(\mU^{T} \vx)$ its \emph{ridge approximation}. Further, we assume that the columns of $\mU$ are orthonormal, i.e., $\mU^T \mU = \mI$. The above definitions imply that the gradient of $f$ is nearly zero along directions that are orthogonal to the subspace of $\mU$. In other words, if we replace $\vx$ with $\vx + \vh$ where $\mU^T \vh = 0$, then $f(\vx + \vh) = g(\mU^T (\vx + \vh)) = f(\vx)$. Visualizing $f_i$ along the coordinates of $\vx_{i}^{T}\mU$ for all designs $i$, can provide extremely powerful inference; such scatter plots are called \emph{sufficient summary plots}.

\subsection{Techniques for dimension reduction}
Techniques for estimating $\mU$ build on ideas from sufficient dimension reduction \cite{cook2009regression} and more recent works such active subspaces \cite{constantine2015active} and polynomial \cite{hokanson2018data, constantine2017near} and Gaussian \cite{seshadri2018dimension} ridge approximations. While our work in this paper is invariant to the specific  parameter-space dimension reduction technique utilized, we briefly detail a few ideas within ridge approximation. Our high-level objective is to solve the optimization problem
\begin{equation}
\underset{\mU\in\mathbb{R}^{d\times m},\; \boldsymbol{\alpha} \in \mathbb{R}^{p} }{minimize}\left\Vert f\left(\vx\right)-g_{\boldsymbol{\alpha}}\left(\mU^{T} \vx\right)\right\Vert _{2}^{2},
\label{equ:ridge}
\end{equation}
over the the space of matrix manifolds $\mU$ and the coefficients (or hyperparameters) $\alpha$ associated with the parametric function $g$. This is a challenging optimization problem and it is not convex. In Seshadri et al.~\cite{seshadri2018dimension} the authors assume that $g$ is the posterior mean of a Gaussian process (GP) and iteratively solve for the hyperparameters associated with the GP, whilst optimizing $\mU$ using a conjugate gradient optimizer on the Stiefel manifold (see Absil et al.~\cite{absil2009optimization}). In Constantine et al.~\cite{constantine2017near} the authors set $g$ to be a polynomial and iteratively solve for its coefficients---using standard least squares regression---whilst optimizing over the Grassman manifold to estimate the subspace $\mU$. It should be noted that these techniques are motivated by the need to break the \emph{curse of dimensionality}. In other words, one would like to estimate both $g$ and $\mU$ for a $d$ dimensional, scalar-valued, function $f$ without requiring a large number of computational simulations. 

The dimension reduction strategy we pursue in this paper is based on active subspaces \cite{constantine2015active} computational heuristic tailored for identifying subspaces that can be used for the approximation in \eqref{equ:ridge}. Broadly speaking, active subspaces requires the approximation of a covariance matrix $\mC \in \mathbb{R}^{d \times d}$
\begin{equation}
    \mC = \int_{\mathcal{X}} \nabla_{x} f \left( \vx \right) \nabla_{x} f \left( \vx \right) ^{T} \boldsymbol{\rho} \left( \vx \right) d\vx, 
\label{equ:active}
\end{equation}
where $\nabla f \left(\vx \right)$ represents the gradient of the function $f$ and $\boldsymbol{\rho}$ is the probability density function that characterizes the input parameter space $\mathcal{X} \in \mathbb{R}^{d}$. The matrix $\mC$ is symmetric positive semi-definite and as a result it admits the eigenvalue decomposition
\begin{equation}
\mC= \mW \boldsymbol{\varLambda} \mW^{T}=\left(\begin{array}{cc}
\mW_{1} & \mW_{2}\end{array}\right)\left(\begin{array}{cc}
\boldsymbol{\Lambda}_{1}\\
 & \boldsymbol{\Lambda}_{2}
\end{array}\right)\left(\begin{array}{cc}
\mW_{1} & \mW_{2}\end{array}\right)^{T},
\end{equation}
where the first $m$ eigenvectors $\mW_{1} \in \mathbb{R}^{d \times m}$, where $m << d$---selected based on the decay of the eigenvalues $\boldsymbol{\Lambda}$---are on average directions along which the function varies more, compared to the directions given by the remaining $\left(d-m \right)$ eigenvectors $\mW_{2}$. Readers will note that the notion of computing eigenvalues and eigenvectors of an assembled covariance matrix is analogous to principal components analysis (PCA). However, in \eqref{equ:active} our covariance matrix is based on the average outer product of the gradient, while in PCA it is simply the average outer product of samples, i.e., $\vx \vx^{T}$. Now, once the subspace $\mW_{1}$ has be identified, one can approximate $f$ via

\begin{align}
\begin{split}
    f \left(\vx \right) = f \left( \mW \mW^{T} \vx \right) &= f \left( \mW_{1} \mW_{1}^{T} \vx + \mW_{2} \mW_{2}^{T} \vx \right) \\
    & \approx g \left( \mW_{1}^{T} \vx \right),
    \end{split}
\end{align}

in other words we project individual samples $\vx_{i}$ onto the subspace $\mW_{1}$. Moreover, as the function (on average) is relatively flat along directions $\mW_{2}$, we can approximate $f$ using the directions encoded in $\mW_{1}$. 

But how do we compute \eqref{equ:active}, as for a given $f$ we may not necessarily have access to its gradients? In \cite{seshadri2018turbomachinery}, the authors construct a global quadratic model to a 3D Reynolds Averaged Navier Stokes (RANS) simulation of a turbomachinery blade and then analytically estimate its gradients. We detail their strategy below as we adopt the same technique for facilitating parameter-space dimension reduction. 

Assume we have $N$ input-output pairs $\left\{\vx_{i}, f_{i} \right\}_{i=1}^{N}$ obtained by running a suitable design of experiment (see \cite{pukelsheim1993optimal}) within our parameter space. We assume that the samples $\vx_{i}  \in \mathbb{R}^{d}$ are independent and identically distributed and that they admit a joint distribution given by $\rho \left(\vx \right)$. Here we will assume that $\rho \left(\vx \right)$ is uniform over the hypercube $\mathcal{X} \in [-1, 1]^{d}$. We fit a global quadratic model to the data, 
\begin{equation}
f(\mathbf{x}) \approx \frac{1}{2}\mathbf{x}^{T} \mathbf{A}\mathbf{x} + \textbf{c}^{T}\mathbf{x} +d,
\label{quadraticeqn}
\end{equation}
using least squares. This yields us values for the coefficients $\mA, \vc$ and the constant $d$. Then, we estimate the covariance matrix in \eqref{equ:active} using
\begin{equation}
    \hat{\mC} = \int_{\mathcal{X}} \left(\mA \vx + \vc \right)\left(\mA \vx + \vc \right)^{T} \boldsymbol{\rho} \left( \vx \right) d\vx.
\end{equation}
Following the computation of the eigenvectors of $\hat{\mC}$, one can then generate sufficient summary plots that are useful for subsequent inference and approximation. 

We apply this quadratic recipe and show the sufficient summary plots for a 3D turbomachinery blade in section \ref{sec:aero}, both in a standard desktop environment and in virtual reality. This comparison---the central objective of this paper---is motivated by the need to explore the gains VR technologies can afford in aerospace design. That said, prior to delving into our chosen case study, an overview of existing immersive visual technologies and their associated frameworks is in order. 



\section{SUPPORTING AEROSPACE DESIGN IN VR} \label{sec:aero}

\label{sec:vr}
\begin{figure}[t!]
    \centering
    \includegraphics[width=\textwidth]{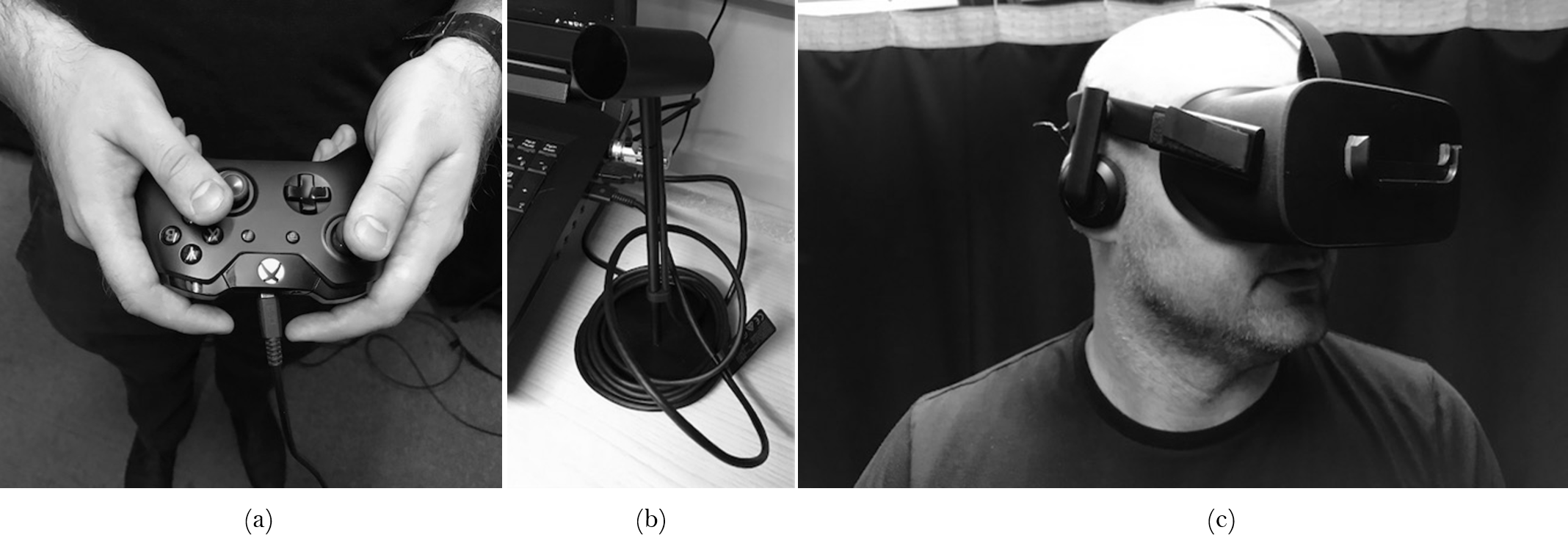}
    \caption{Hardware interfaces in VR: (a) Xbox controller; (b) Oculus Rift's motion sensor; (c) Oculus Rift VR headset.}
    \label{fig:controller}
\end{figure}

Visual analytics (VA), a phrase first coined by J.J.~Thomas et al.~\cite{Thomas:2006:VAA:1110637.1110648} has two ingredients: 1) an interactive visual interface \cite{Thomas:2006:VAA:1110637.1110648}; and 2) analytical reasoning \cite{Thomas:2006:VAA:1110637.1110648}. Recent advances and breakthroughs in the development of VR (virtual reality) and AR (augmented reality) have spun off another branch of research known as \textit{Immersive Analytics}~\cite{immersive_analytics_2015} (IA). IA seeks to understand how the latest wave of immersive technologies can be leveraged to create more compelling, more intuitive and more effective visual analytics frameworks.
There are still numerous hurdles to overcome for VR-based data analytics tools to be widespread. Issues associated with any type of a 3D interface, for example, potential occlusion effects \cite{Shneiderman:2003:WMI:950627.950723}, high computing power demands and specialized, (often costly) hardware, have to be resolved, or at least minimized. Moreover, interaction techniques have to facilitate a user's understanding of the visualization and avoid becoming a distraction. Finally, certain guidelines have to be incorporated to mitigate the risk of the simulation sickness \cite{kennedy_sim_sickness, ruddle_sim_sickness} symptoms that can manifest during or immediately after the use of a VR headset.

Many interaction techniques and devices have been conceived that can be used separately or in combination for interaction and control of a VR environment. In this paper we use the standard off-the-shelf Xbox \cite{xbox} controller shown in Fig.~\ref{fig:controller}(a) that comes prepacked with the Oculus Rift\cite{oculus} bundle (see Fig.~\ref{fig:controller}(b) and Fig.~\ref{fig:controller}(c)). This controller-style has achieved very high adoption in gaming industry; its design is ergonomic and easy to learn. As one example of wider adoption, the US Navy recently adopted the use of a Xbox controller to operate the periscope on nuclear-powered submarine\footnote{\url{https://www.digitaltrends.com/cool-tech/navy-xbox-controllers-attack-submarines/}, Last accessed: August 2019}. 



Here, we present a VR aerospace design environment with a focus on dimension reduction. Information on one of the earliest examples of research into using VR for applications in the scope of aerospace design can be found in Hale~\cite{hale_applied_1994}. Garc\'ia-Hern\'andez et al.~\cite{garcia-hernandez_perspectives_2016} pointed out that the VR technology is starting to gain ground in aerospace design and listed a range of aerospace research topics in which VR already is, or can be, successfully applied. This includes, among others, \textit{spacecraft design optimization} (e.g. Mizell~\cite{mizell_virtual_1994} discusses use of VR and AR in aircraft design and manufacturing whereas Stump et al.~\cite{stump_trade_2004} used IA to aid a satellite design process) and \textit{aerodynamic design}, in which 3D scatter plots are already in-use (see Jeong et al.~\cite{jeong_data_nodate}). Other applications include use of the haptic feedback (e.g. Savall et al.~\cite{savall_description_2002} describes REVIMA system for maintainability simulation and Sagardia et al.\cite{sagardia_interactive_2013} presents the VR-based system for on-orbit servicing simulation), collaborative environments (e.g. Roberts et al.~\cite{roberts_collaborative_2015} introduces an environment for the Space operation and science whereas Clergeaud et al.~\cite{clergeaud_3d_2016} discusses implementation of the IA tools used in context of the aerospace with Airbus Group), aerospace simulation (e.g. Stone et.al~\cite{stone_evolution_2011} discuss the evolution of aerospace simulation that uses immersive technologies), telemetry and sensor data visualization (e.g. see Wright et al.~\cite{wright_immersive_2001}, Lecakes et al.~\cite{lecakes_visualization_2009} or Russell et al.~\cite{russell_acquisition_2009}), or planetary exploration (e.g. see Wright et al.~\cite{wright_immersive_2001}). Garc\'ia-Hern\'andez et al.~\cite{garcia-hernandez_perspectives_2016} suggests that three elements are especially promising for a VR-based approach: 1) \textit{integration of multiple 2D graphs for 3D data} \cite{garcia-hernandez_perspectives_2016}; 2) \textit{3D parallel coordinates} \cite{garcia-hernandez_perspectives_2016} (see Tadeja et al.~\cite{tadeja_chi_2019, tadeja_aiaa_2020}); and 3) \textit{visualization of complex graphs} \cite{garcia-hernandez_perspectives_2016}.
In this paper, we loosely follow (1), but with a key difference: we use subspace-based dimension reduction to generate the 3D graphs for high-dimensional data.

\subsection{Applications in design}
\begin{figure}[t]
    \centering
    \begin{subfigmatrix}{2}
        \subfigure[]{\includegraphics{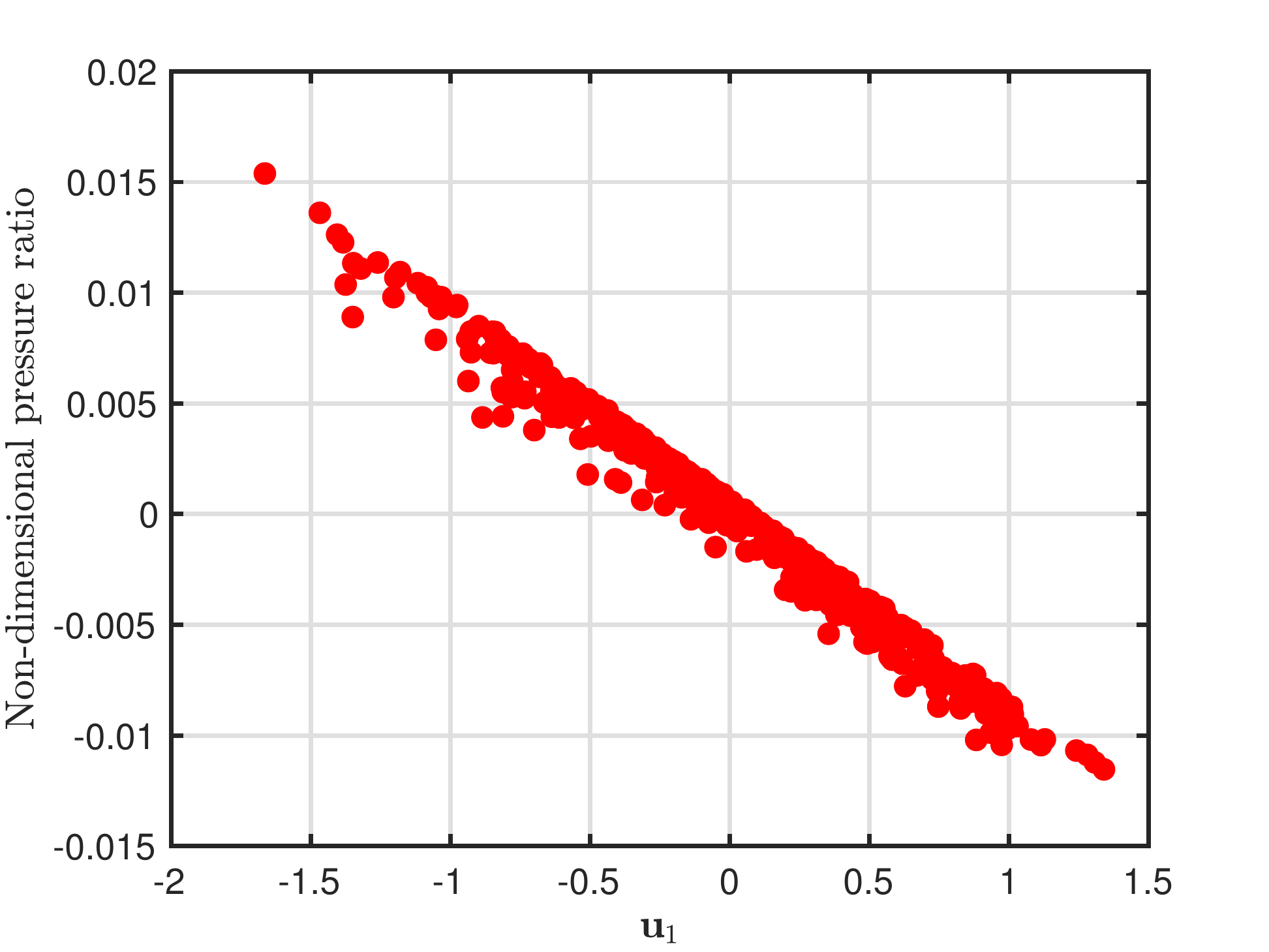}}
        \subfigure[]{\includegraphics{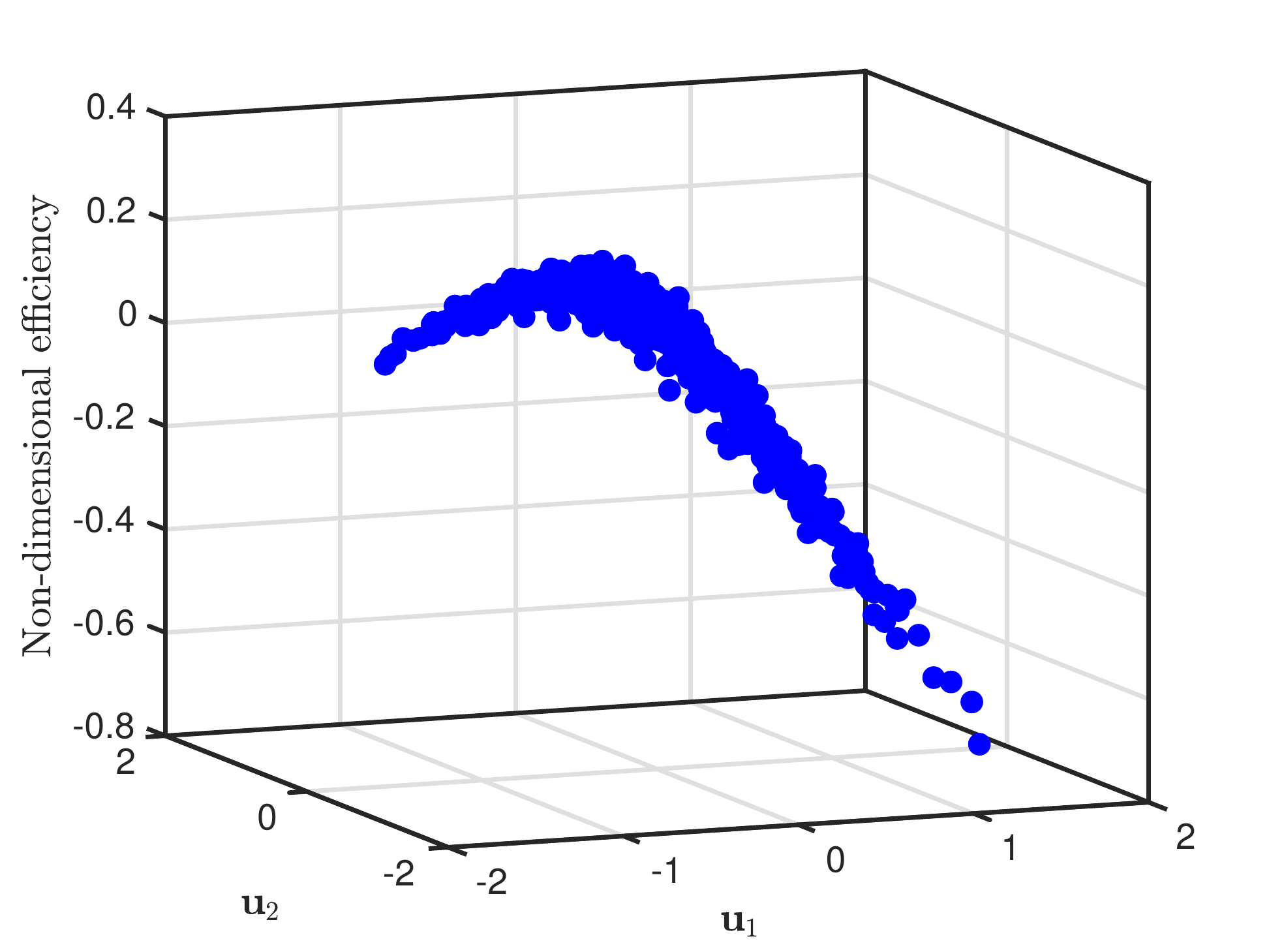}}
    \end{subfigmatrix}
    \caption{Sufficient summary plots of (a) Pressure ratios; (b) Efficiency, for a range of different computational designs for turbomachinery blade, obtained from a design of experiment study. Based on work in Seshadri et al.~\cite{seshadri2018turbomachinery}.}
    \label{fig:TURBO}
\end{figure}

Our dimension reduction results and case study is based on the work undertaken in Seshadri et al.~\cite{seshadri2018turbomachinery}. Here the authors study the 25D design space of a fan blade using the quadratic active subspaces recipe detailed in \ref{sec:dim}. Towards this end, we used the design of an experiment with $N=548$ 3D RANS computations with different designs; the design space used in this study included five degrees of freedom specified at five spanwise locations. These degrees of freedom comprised of an axial displacement, a tangential displacement, a rotation about the blade's centroidal axis, leading edge recambering and trailing edge recambering, specificed at $0, 25, 50, 75$ and $100 \%$ span. Thus, we obtained values of the efficiency and pressure ratio for each design vector $\vx_{i}$. These are two important output quantities of interest in the design of a blade. By studying the eigenvalues and eigenvectors of the covariance matrix for these two objectives the authors were able to discover a 1D ridge approximation for the pressure ratio of a fan and a 2D ridge approximation for the efficiency. These sufficient summary plots are shown in Fig.~\ref{fig:TURBO}. There are a few important remarks to make regarding these plots.

For the pressure ratio sufficient summary plot, shown in Fig.~\ref{fig:TURBO}(a), the horizontal axis is the first eigenvector of the covariance matrix associated with the pressure ratio, $\vu_{1}$. For the efficiency sufficient summary plot, shown in Fig.~\ref{fig:TURBO}(b), the two horizontal axes are the first two eigenvectors of the covariance matrix associated with the efficiency $[\vu_{1}, \vu_{2}]$. It is important to note that the subspaces associated with efficiency and pressure ratio are distinct.

\begin{figure}[t!]
    \centering
    \includegraphics[width=\textwidth]{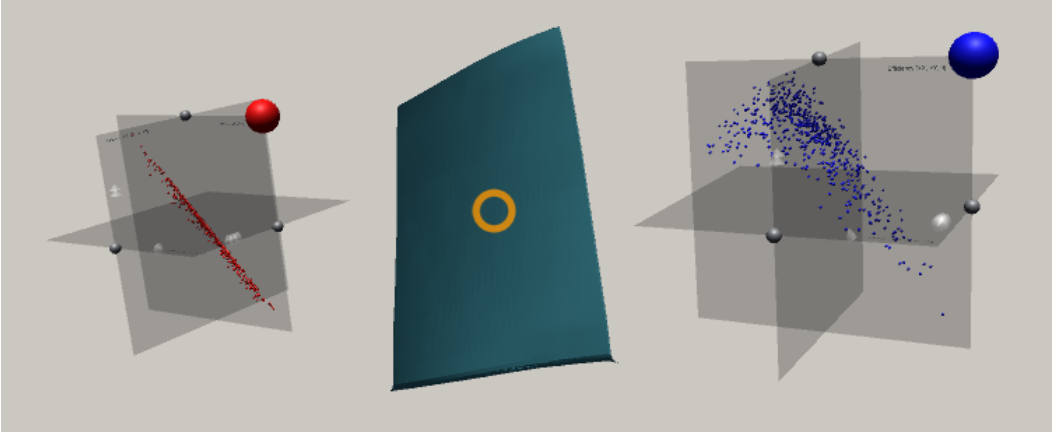}
    \caption{The user's field-of-view: the right-hand plot can, for example, show the lift coefficients whereas the left-hand side plot can contain the drag coefficient values. The nominal blade geometry is visualized in the middle between the two plots. The orange circle is a cross-hair singalising where user is looking at the moment.}
    \label{fig:field_of_view}
\end{figure}

The sufficient summary plots above permit us to identify and visualize low-dimensional structure in the high-dimensional data. More specifically, these plots can be used in the design process as they permit engineers to make the following inquiries:

\begin{itemize}
    \item What linear combination of design variables is the most important for increasing / decreasing the pressure ratio?
    \item How do we increase the efficiency?
    \item What are the characteristics of designs that satisfy a certain pressure ratio?
    \item What are the characteristics of designs that satisfy the same efficiency?
\end{itemize}

We use these sufficient summary plots in a bespoke VR environment. Our high-level objective is to ascertain whether it is possible to leverage tools in VR in conjunction with parameter-space dimension reduction to facilitate better design decision-making and inference. To achieve this goal, we seamlessly integrate the aforementioned sufficient summary plots with the 3D geometric design of the blade, i.e., 

\begin{equation*}
    \text{pressure ratio} \; \; \rightleftharpoons \; \; \text{geometry visualization} \; \;  \rightleftharpoons  \; \; \text{efficiency}.
\end{equation*}

In other words, as the user selects a different design---by selecting a suitable level of performance from the sufficient summary plots---they visualize the geometry of the blade that yields that performance. Moreover, they should be able to compare this geometry with that of the nominal design. We clarify and make precise these notions in the forthcoming subsections. 

\subsection{Function structures}
\begin{figure}[t!]
    \centering
    \includegraphics[width=\textwidth]{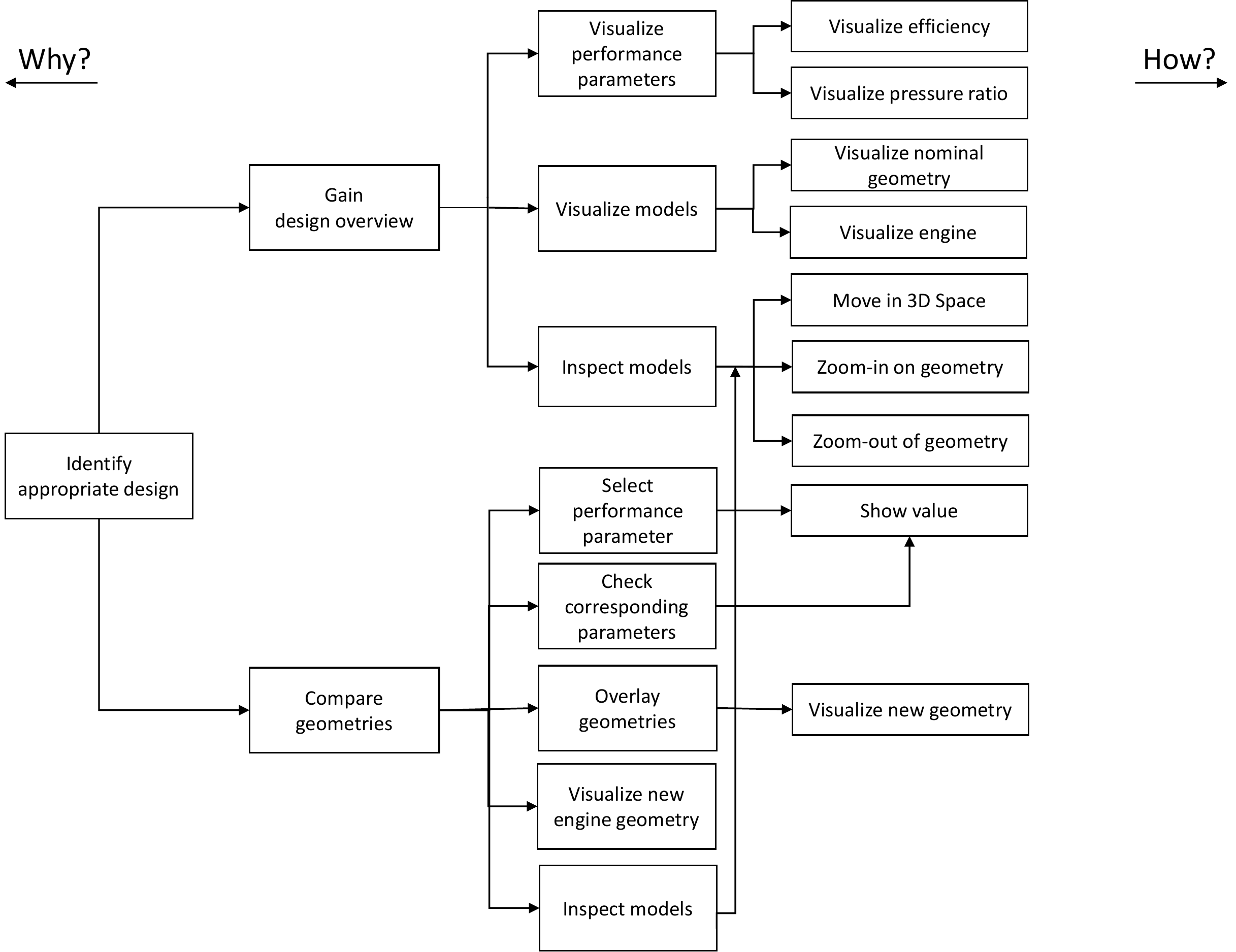}
    \caption{A functional model of the interactive system.}
    \label{fig:fast_diagram}
\end{figure}

We model the function structures of the system using \textit{Function Analysis Systems Technique}~\cite{Shefelbine_edc} (FAST). Fig.~\ref{fig:fast_diagram} shows the function structures of the VR visualization environment for an aerospace design workflow with dimension reduction. The FAST-diagram in Fig.~\ref{fig:fast_diagram} models the level of abstraction on the horizontal axis and function sequence on the vertical axis.

\subsection{System structure}
\begin{figure}[h!]
    \centering
    \includegraphics[width=1\textwidth]{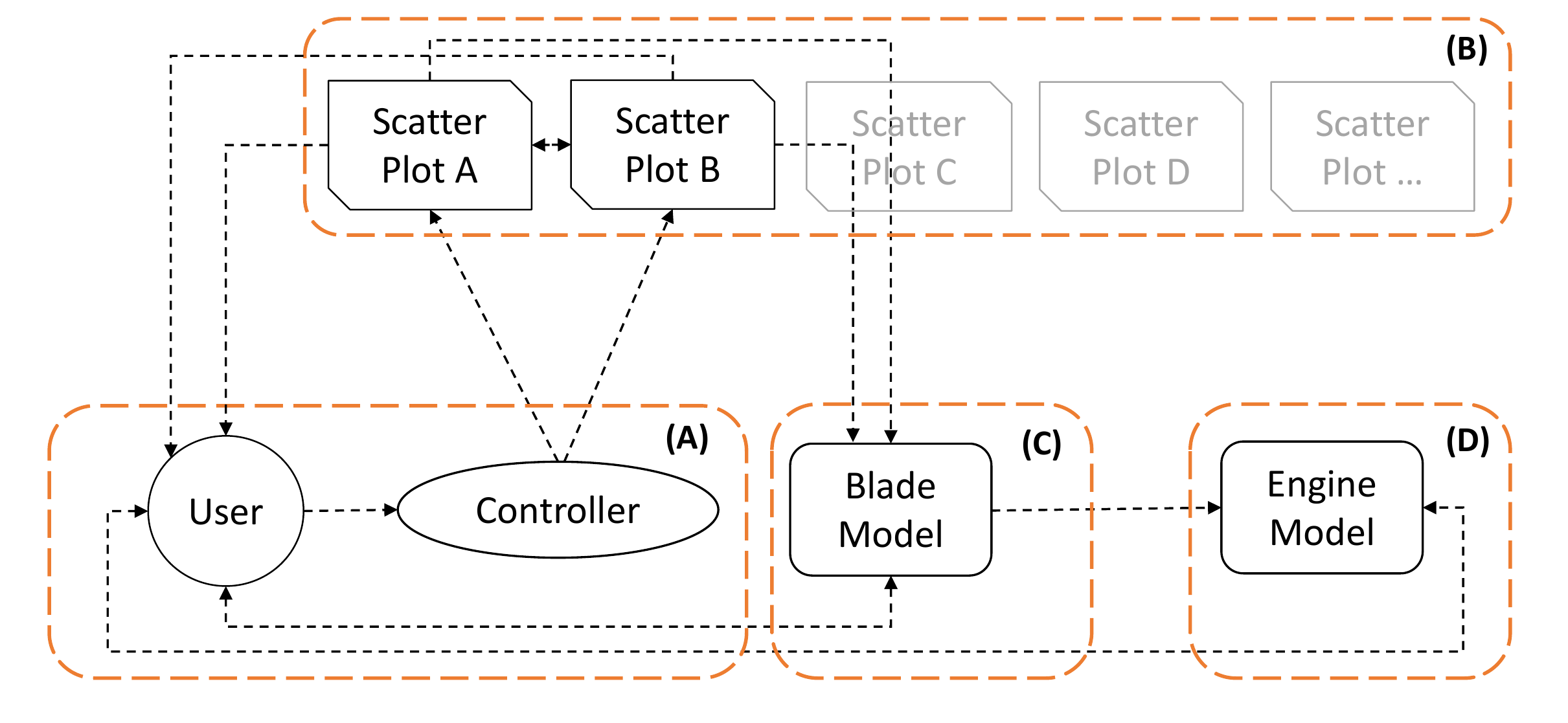}
    \caption{The diagram shows how the signals are flowing within the system between its four main components:  (A) the user grouped together with a controller used for user input; (B) a set of performance parameters visualized as the 3D scatter plots, in this case, efficiency and pressure-ratio 3D scatter plots; (C) blade geometry model; and (D) complete engine geometry model.}
    \label{fig:signals}
\end{figure}

We model the system internal structure by observing the internal flow of signals between the individual system elements. The visualization consists of four distinguishable parts: (A) the user who is responsible for all the actions of the system once the data had been loaded and visualized, (B) efficiency and pressure-ratio 3D scatter plots, (C) blade model, and (D) engine geometry model. The signals are usually bi-directional and can introduce a \emph{chaining effect}. For instance, when user is gazing over an interactive object, which is internally facilitated by the ray-tracing, the object highlights itself, that is, the user receives a return feedback signal in the form of a visual clue. Moreover, selection of a data point on the scatter plot through an implementation of the \textit{linking \& brushing} interaction technique, leads to a selection of the mapped data point on the other scatter plot and visualization of a new geometry overlapping with the nominal shape. The signal flow analysis is presented on Fig.~\ref{fig:signals}. The main signal flows are decomposed into:

\begin{enumerate}
\item \textbf{The user:} The user interacts with the system using a combination of gaze-tracking and ray-tracing. This works as follows. Gaze-tracking is achieved with the help of a cross-hair in the middle of user’s field of view, placed a certain, fixed distance along the camera’s forward direction. Rays extending from the cross-hair are constantly checked for intersection with other interactive objects i.e. data points on the scatter plots. If such an interaction occurs, the object automatically highlights, providing a signal to the user that it can be interacted with. The way in which the user directly receives signals from other parts of the visualization is unidirectional, that is, a user's action results in a visual response. The way the user interacts with other objects is through a combination of gaze-tracking and ray-tracing (i.e., an orange cross-hair, see Fig.~\ref{fig:field_of_view} and Fig.~\ref{fig:blade_vr}) as well as actions invoked with a tap of a button (see Fig.~\ref{fig:controller}).
\item \textbf{3D scatter plots:} The scatter plots receives signals from the user through a mixture of gaze-tracking and ray-tracing inputs combined with the tap of a button on the controller. This is reflected back to the user by, for example, highlighting scatter plots elements, such as data points or movement selectors, that are being gazed over or changed their color after selection. In turn, this action invokes unidirectional changes in the visualized blade geometry and the turbofan engine.
\item \textbf{Blade geometry model:} The blade geometry visualization receives signals from both scatter plots by the user performing a selection of a data point on any of the plots, which automatically visualizes the new blade geometry. Moreover, even though the user cannot directly influence the geometry, by using the movement and maneuvering techniques in the system, the user can inspect the geometry by zooming in on its internal and external surfaces. Hence the relation between the scatter plots and the blade is unidirectional, whereas the relation between the user and the blade model is bidirectional. Furthermore, once the new blade has been visualized, the visualization of the hub with blades in the engine model simultaneously changes as well. This can be thought of as another unidirectional relation as it cannot happen the other way around.
\item \textbf{Engine geometry model:} Once the new geometry shape is selected by the user, the blade row with the series of blades embedded in the engine model is automatically replaced. This change is immediately visible to the user providing visual feedback.
\end{enumerate}

\subsection{Tasks analysis}
From  the  limitations imposed by the current state and understanding of the VR environment and from our own analysis of the system achieved by the FAST analysis (see Fig.~\ref{fig:fast_diagram}) we identified two primary, high-level tasks:

\textbf{T1---Gaining design overview}: The system should permit the user to easily gain an overview of the entire design space i.e. possible blades geometries together with their associated performance parameters.

\textbf{T2---Compare geometries}: The system should permit the user to easily compare the nominal blade geometry with the one associated with a particular set of performance.

\noindent These two main tasks (i.e. \textit{T1} and \textit{T2}) were supported and augmented by a number of low-level tasks:

\textbf{T3---Movement and interaction}: Due to the nature of spatial, 3D immersive workspace provided by the VR environment, this task has a dominant and a supporting role with respect to all the other tasks. Movement, maneuvering and interaction are achieved through the gaze-tracking and with the help of a gamepad controller. All movement facilitated by either the joystick [\textit{J}] or triggers [\textit{T}] (see Fig.~\ref{fig:xbox_controller}) takes place with respect to the users gaze (see orange cross-hair on Fig.~\ref{fig:field_of_view}). Moreover, the user can interact with an object through gazing over an object and tap of a button (see Fig.~\ref{fig:xbox_controller}). Zooming in or out on a part of the visualization is also achievable by the user's movement in the virtual space.

\textbf{T4---Visualization of performance parameters}: The performance parameters, such as efficiency and pressure ratio that were used in our case, are visualized as an interactive 3D scatter plots floating in the 3D space. These can be freely moved, rotated about each of the main axes and implements the \textit{linking \& brushing} interaction technique i.e. changes in one scatter plot are simultaneously reflected on the other scatter plot and blade and engine visualizations as well. This task mainly supports \textit{T1}.

\textbf{T5---Visualization of blades and engine models}: The nominal blade geometry and associated engine visualization are immediately visible at the start of the visualization. Once the new geometry is selected, the nominal blade renders semi-transparent and the shape of the new blade is superimposed over it. Moreover, the hub with a row of blades are substituted with the new geometries in the engine model.

\textbf{T6---Models inspection}: The inspection of the changes in the engine visualization and the blade geometry itself can be made through the \textit{T3} task.

\subsection{Visualization framework}
The visualization framework is built using Unity3D---one of the most widely used game engines with built-in VR development support. Both of the two mainstream VR headsets provide supporting packages developed natively for Unity3D, which substantially speeds up the development process. This software is built on top of the \textit{Unity VR Samples pack}\cite{unity_vr_samples_pack} and uses the \textit{Oculus Utilities for Unity}\cite{oculus_utilities} package as well as parts of the Unity asset\cite{unity_off_screen_arrows}. In addition, we use the asset store available for the Unity3D game engine, which contains many VR-ready tools and supporting packages. 

A survey by Wagner et al.~\cite{wagner_native_2016} highlights that game engines ``\textit{do not support any data exploration}''\cite{wagner_native_2016} techniques. In other words, these features have to be designed and implemented from scratch. To allow user interaction with data we use an Xbox Controller connected with the laptop via USB cable (see  Fig.~\ref{fig:controller}(a)) in combination with gaze-tracking through a cross-hair which moves with the user's head and is placed straight from the camera (visualized as an orange cross-hair, see Fig.~\ref{fig:field_of_view}, Fig.~\ref{fig:blade_vr} and Fig.~\ref{fig:scatter_plot}). 



\subsection{Interaction and movement}
\begin{figure}[ht]
    \centering
    \includegraphics[width=1\columnwidth]{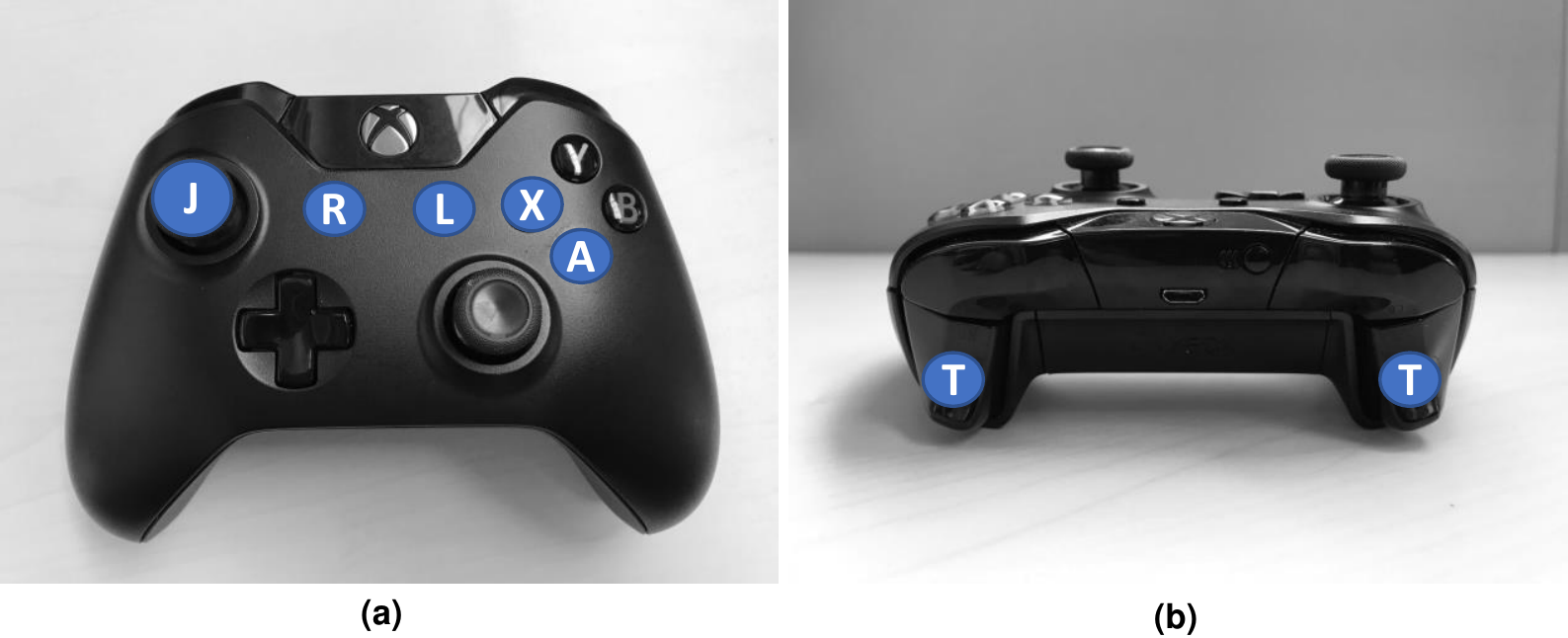}
    
    \caption{The Xbox controller: (a) shows the top view with the left-hand joystick $[J]$ used to control the 2D movement on the $X$-$Z$ plane whereas (b) shows the front view with the two triggers $[T]$ responsible for vertical movement along the $Y$ axis. The other action buttons indicated in (a) have the following meanings: $[R]$ for reload of the visualization; $[L]$ for loading next dataset; $[A]$ for selection of an interactive item; and $[X]$ for moving or rotating the scatter plots.}
    \label{fig:xbox_controller}
\end{figure}

The interaction is designed around gaze-tracking in combination with the standard buttons on the Xbox\cite{xbox} controller (see Fig.~\ref{fig:xbox_controller}).
Supported interactions are mapped as follows:
\begin{itemize}
    \item Left-hand joystick and $[T]$ buttons (see Fig.~\ref{fig:xbox_controller}): Triggers movement along the $X$-$Z$ plane and movement along the vertical axis respectively, right-hand $[T]$ is assigned to ``up'' and left-hand $[T]$ is ``down''. The movement in the $X$-$Z$ plane is always with respect to the user's gaze. This manoeuvring combination permits the user to move in any direction and in any position in 3D space. The user moves with constant velocity and with fluid movement to ensure the user is receiving continuous closed-loop visual feedback on their changing position in relation to the surroundings.
    \item Action button $[A]$: Selects an interactive element, such as a scatter plot rotation and movement selector, or a data point (see Fig.~\ref{fig:scatter_plot}). Objects highlight themselves when the user's gaze, as indicated by a cross-hair, is on them. Double-tapping on the $[A]$ button selects the highlighted object.
    \item Button $[X]$: When tapped after the selection of a scatter plot point, it will re-position the point to a certain distance towards the user's present gaze direction. Furthermore, if the rotation selector is active, selecting this button will initiate the scatter plot's rotation over $90^{\circ}$ based on the current direction of the user's gaze.
    \item Button $[R]$: Resets the visualization and all its associated elements to their original state.
    \item Button $[L]$: This button loads the next dataset: a new set of performance parameters and associated blade geometries.
\end{itemize}


\vspace{-4mm}
\subsection{Blade and engine visualizations}
\begin{figure}[t!]
    \centering
    \includegraphics[width=\textwidth]{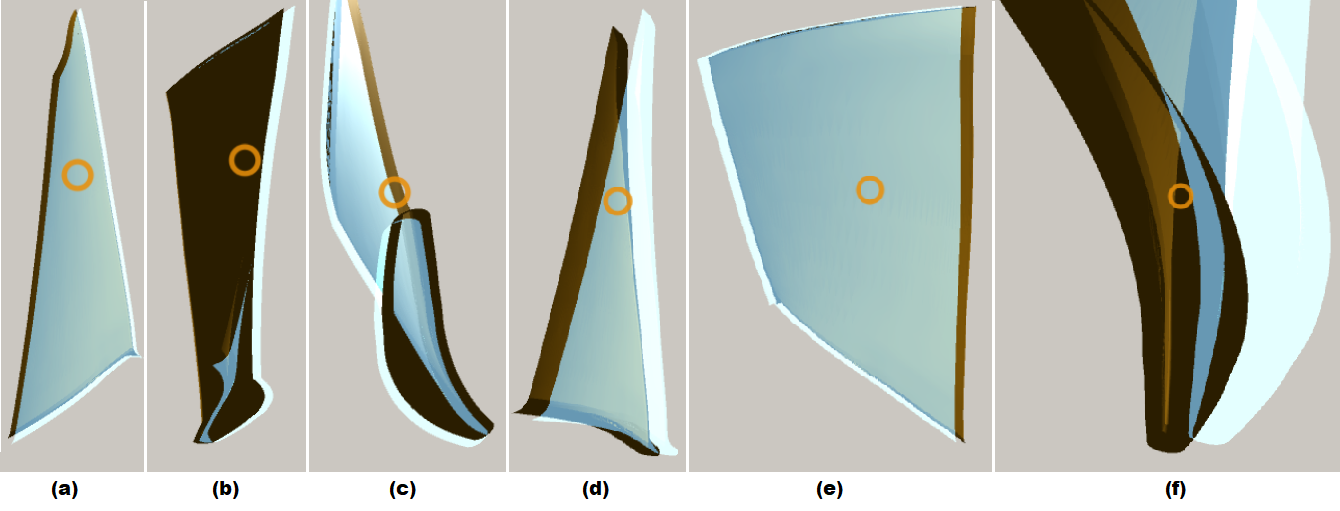}
    \caption{By selecting any data point on any plot, the user can immediately observe the blade's geometry associated with this particular design. Moreover, the user can observe and compare the differences between the nominal and perturbed geometries as the former is kept rendered as a semi-transparent shape overlaying the latter. As users can freely maneuver in 3D space they can visually inspect the entire blade from any direction and zoom in on any of its parts, as shown in (a--f).}
    \label{fig:blade_vr}
\end{figure}

As alluded to previously, the central artifact in our VR environment is the geometry of the designs. Our virtual environment contains as many geometries as there are data points, resulting in a total of 548 stereo lithography (STL) files. Hence, whenever a data point is selected on one of the plots, the accompanying shape is instantly visualized. To provide the user with a quick and an effective way of comparing the new perturbed design, the nominal geometry is still kept visible and rendered as a translucent object, as can be seen in Fig.~\ref{fig:blade_vr}. This solution, combined with unlimited movement dexterity, allows the user to visually inspect and observe any differences between the two geometries. Furthermore, by simply changing their position, or by tilting their head (thereby changing the rotational angle), the user can zoom-in and zoom-out on any of the blades parts for a close inspection as presented in Fig.~\ref{fig:blade_vr}(f).

\begin{figure}[!ht]
    \centering
    \includegraphics[width=\columnwidth]{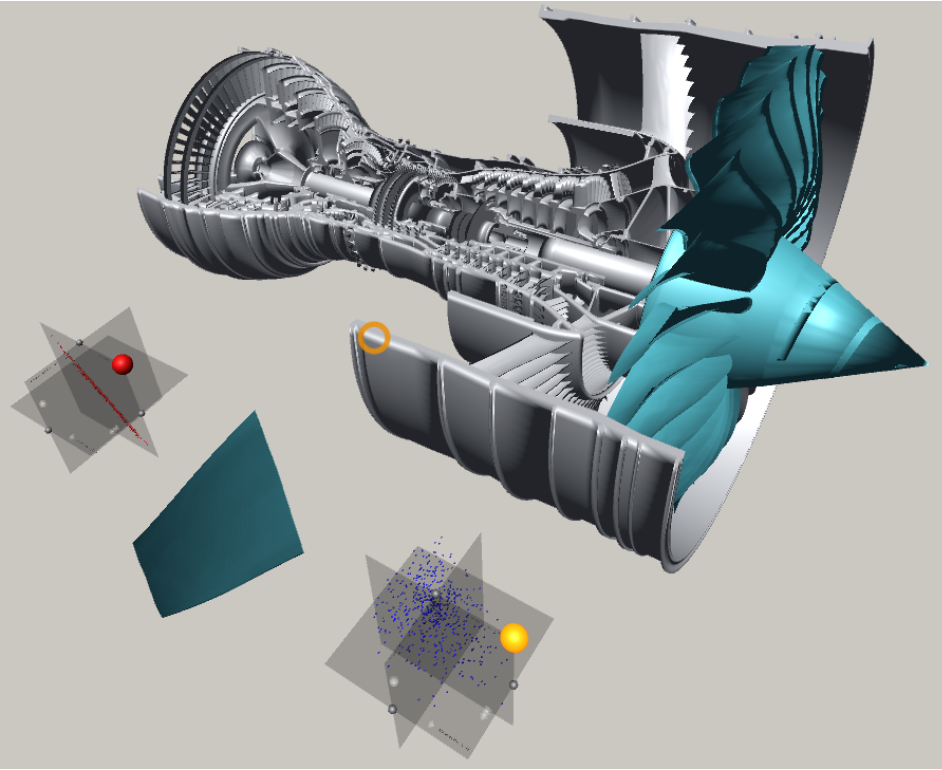}
    \caption{The entire visualization as it is seen by the user with the complete engine model in the back and the two 3D scatter plots and the nominal blade geometry (in blue) in front. The hub with a series of blades is also shown (in blue).}
    \label{fig:full_engine_A}
\end{figure}

\begin{figure}[!ht]
    \centering
    \includegraphics[width=\columnwidth]{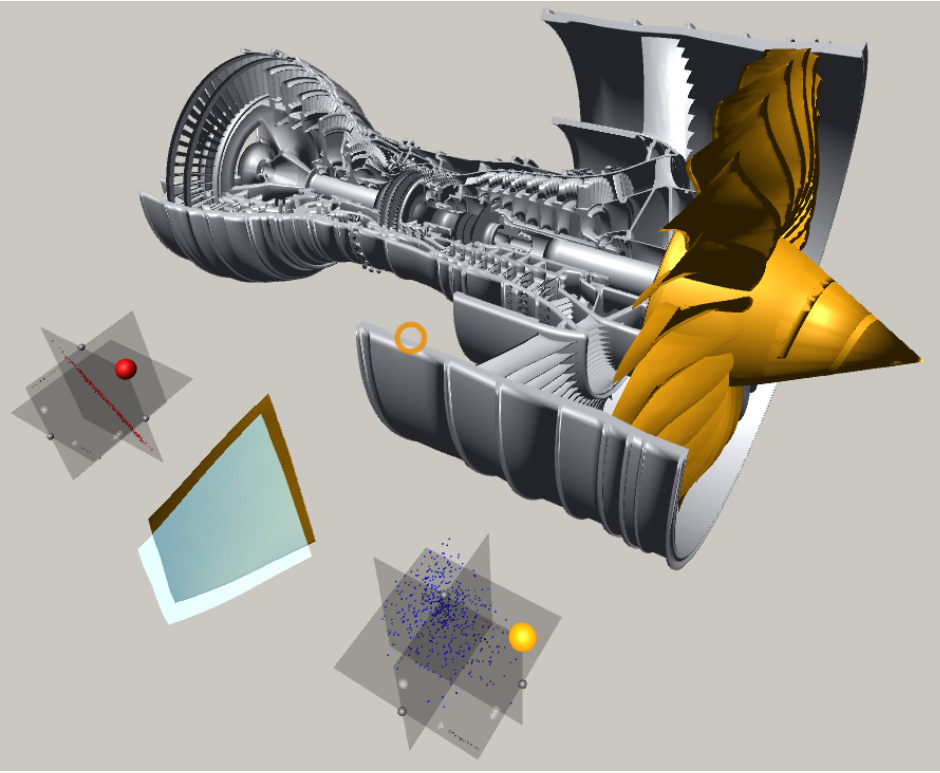}
    \caption{The same view as in Fig.~\ref{fig:full_engine_A} with a single data point selected on one of the scatter plots (the one on the bottom-right). The nominal blade geometry was rendered as semi-transparent shape with a new geometry superimposed on top of it. The engine hub with a new series of blades is also shown (in tan).}
    \label{fig:full_engine_B}
\end{figure}


The visualized engine model~\cite{shakal} (see Fig.~\ref{fig:full_engine_A}) consists of six independent parts including the hub with connected blades. When a new blade geometry is being investigated by the user, the blades visible in the engine are automatically replaced as well. Due to limitations imposed by the used CAD model itself, it was not possible to substitute the blades individually, thus the entire hub with the series of attached blades could be replaced altogether which is signaled to the user by changing color of this entire part (see Fig.~\ref{fig:full_engine_B}).



\subsection{Sufficient summary plots}
The key element in the framework is the sufficient summary plots; they are visualized as three fixed-size, axis-aligned translucent orthogonal rectangles. The data points are scaled so the values of their respective coordinates are within the range of the translucent surfaces. When any of the spheres denoting a data point is selected, the marker lights up and switches to a selection color (light green). Moreover, as we have a 1:1 mapping between the plots, the corresponding design on the other plot is selected. Furthermore, a number of semitransparent cones\footnote{W. Kresse, used under CC BY-SA 3.0;  \url{http://wiki.unity3d.com/}} was embedded into these sufficient summary plots to denote their axes: one for the $X$-axis, two for the $Y$-axis and three for the $Z$-axis. Selection of a shape with its pointing tip has an additional advantage---the orientation informs the user of the positive side of a given axis. Each selection can be reverted by double-tapping the $[A]$ button while gazing over it.

The 3D spheres in the plots were used to denote both the data points and various selectors' markers. Using shape perception has a long-standing application history for VR-based visualizations. For instance, Ribarsky et al.~\cite{ribarsky_visualization_1994} used simple 3D shapes such as cones, spheres and cuboids in their system. They also highlight that glyphs with their intrinsic characteristics, such as \textit{``position, shape, color, orientation and so forth''}~\cite{ribarsky_visualization_1994} are very useful when visualizing complex datasets. 

\subsubsection{Initial placement and re-positioning}
\begin{figure}[t!]
    \includegraphics[width=\textwidth]{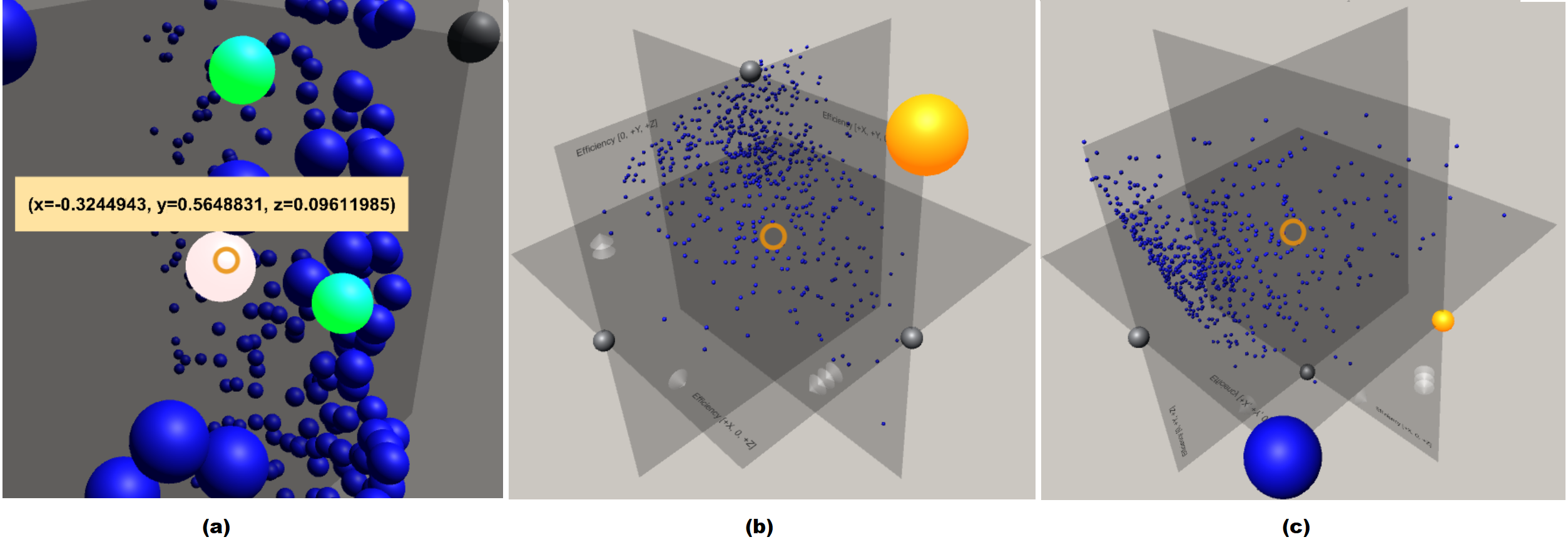}
    \caption{The scatter plot as it is seen by the user. Figure (a) shows a user's gaze (orange cross-hair) hovering over a data point which instantly displays the associated values (e.g. its coordinates) in. Visible, formerly selected points (in green) are reflected on the other scatter plot and the associated geometries are also shown. Figure (b) shows the same plot from a distance. The highlighted spheres (in orange) are the movement selectors: if the user gazes at any point in space and taps the $[X]$ button on the controller the plots will be translated towards that point in space. Figure (c) shows a rotation by $90^{\circ}$ towards the user along the $X$-axis with the axis rotation selector highlighted (in orange). Only a single rotation selector can be active at once across all the scatter plots.}
    \label{fig:scatter_plot}
\end{figure}

The two scatter plots---one for pressure ratio and another for efficiency---are automatically positioned on both sides of the blade, which in turn is positioned in front of the initial user's field of view; see Fig.~\ref{fig:field_of_view}. The plots are placed at the same, pre-configured distance from the user, at a roughly $45^{\circ}$ angle from the $X$-axis.

To ensure that the user does not feel constrained in a nearby region and to make better use of virtually infinite 3D space provided by the VR-environment, users are provided with the possibility of moving the scatter plots. The interaction occurs via gaze-tracking and the \textit{select \& move} metaphor. Every scatter plot has a color-coded interactive sphere, that is, a selector (see Fig.~\ref{fig:scatter_plot}(b)), attached to it in the right-hand top corner. When the user's gaze hovers over it, the selector automatically highlights it and, if selected by double-tapping the $[B]$ button, changes its color to orange (see Fig.~\ref{fig:scatter_plot}(b)). If the user presses the $[X]$ button while a selection is active, the scatter plot is re-positioned at a certain distance towards the point determined by the user's current gaze. If the button is held the plot will follow the cross-hair's movement. 

It is also possible to move both plots at once if more selectors are simultaneously active. In such a scenario, to keep the current relative position of these scatter plots, a barycentre $B=(x_B, y_B, z_B)$ of all these objects is calculated using the formula:

\begin{equation}
B = \frac{1}{N}(x_B=\sum_{i=0}^{N} x_i, y_B=\sum_{i=0}^{N} y_i, z_B=\sum_{i=0}^{N} z_i).    
\end{equation}


\noindent This point is moved along the forward vector from the camera in the same manner as in the case of a single scatter plot. Selected objects are then grouped together and displaced with respect to the new position of the barycentre whilst simultaneously keeping their internal (current position with respect to the local axes) and external (axis-alignment of surfaces) rotations.


\subsubsection{Scatter plot rotation}
The scatter plot can be rotated in $90^{\circ}$ steps about one of the three main axes. This is achieved by double-tapping the $[A]$ button while gazing over one of the three rotation selectors (see Fig.~\ref{fig:scatter_plot} (b-c)). A further press of the $[X]$ button will result in a rotation of all the plot's components, including data points, rotation and movement selectors and axis cone markers. If users are situated in such a way that their gaze is located exactly in front of the active axis selector, the rotation will occur towards the direction provided by the camera's forward vector. Similar to the movement, the plot's orientation in the global coordinate system will not be affected.

\subsubsection{Relationships between the visualization elements}
The data points on one of the plots are correlated in a 1:1:1 (one-to-one-to-one) manner onto the other plot and vice-versa. Moreover, each of the data points is mapped onto one-and-only-one unique blade design. Therefore, whenever a marker is selected on one of the plots, the system will automatically highlight the corresponding data point on the other plot and switch the visualized blade onto the new, corresponding shape. Furthermore, the nominal shape will be kept as a translucent point of reference (see Fig.~\ref{fig:blade_vr}) that overlays the new design to show the user how, where, and to what degree, the new shape differs from the nominal one.

\subsubsection{Labeling}
Both the data points and the axes are automatically labeled. In case of the latter, the strings embedded into the edges of the semitransparent rectangles denoting the axes are read directly from the input text file (see Fig.~\ref{fig:scatter_plot} (a)). The labeling of the data points is only visible once the user is hovering with his or her gaze over the marker (for example, a sphere) and disappears once the user looks at another point, or other parts of the visualization (see Fig.~\ref{fig:scatter_plot} (a)). Furthermore, the small box with the values associated with the point (in this case its coordinates) is always rotated towards the user and follows their gaze. It is rendered on top of any other visualization elements as seen in Fig.~\ref{fig:scatter_plot} (a).


\section{INTERFACE VERIFICATION} \label{sec:eval}
The VR aerospace design environment in this paper is still at an early stage and the objective of this paper is not to present a complete solution but to demonstrate potential benefits of VR for aerospace design.

Here, we verify the fundamental usability of the system using two formative evaluation methods. First, we assess the usability of the system using Nielsen's \cite{nielsen_usability_1994, nielsen_enhancing_1994} guidelines. Second, we use the cognitive dimensions of notations framework \cite{green_cognitive_1989, green_cognitive_1990} to reason about the expressiveness of the system.

\subsection{Usability}
\begin{itemize}
\item \textit{Visibility of system status}: The system provides immediate feedback to users in response to their actions. For instance, whenever a user's gaze hovers over an interactive object (such as, for example, movement and rotation selectors or a data point) it is instantly highlighted. In addition, once an object is selected, the object also changes its color in response. In addition, following the selection of a data point, its corresponding data points on the other plot are also simultaneously selected and the accompanying geometry is loaded automatically. This ensures the user remains synchronized with the system's current state. Finally, whenever the user's gaze is hovering over an object, a gaze-locked text is also displayed to the user, which reveals values associated with this particular data point.

\item \textit{Match between the system and the real-world}: First, the system uses the cross-hair concept which is well known in the real-world to focus and help guiding the users' gaze on objects placed directly behind or near it. Moreover, the 3D scatter plots were designed to immediately resemble their two- or three-dimensional desktop-based counterparts. In addition, initially a user observes all the main elements of the visualization in the field-of-view placed at roughly the same height and direction as how they would be perceived in the real-world if they were visualized on standard computer displays.

\item \textit{User control and freedom}: The user can either load the new set of data or reload the entire visualization with a click of a button. No direct ``undo'' and ``redo'' actions were directly implemented, however, users can always deselect any object or redo the last executed operation. Moreover, using a combination of the gaze-tracking and controller-based interaction, users can locate themselves at any position in 3D space.

\item \textit{Consistency and standards} and \textit{Flexibility and efficiency of use}: As the VR in its current, almost fully immersive form, is a fairly recent development, the technology itself, not to mention its main applicability areas or interaction design principles, is not yet fully understood. However, the system design is as consistent as possible, for instance, all interactive objects can be (de)selected using exactly the same method.


\item \textit{Error prevention}: Measures to prevent the user from errors, such as missing or broken input data (for example, geometries and data points), are  directly built into the system. As there is a 1:1:1 mapping between the system elements (data points) on the two plots and the geometries, missing any of the elements would lead to omitting this particular entry from the visualization and detailed information of such event being written into the log file. Hence, the main error-prone conditions are eliminated. In addition, the system incorporates certain constraints, such as a user cannot have more than a single axis-rotation selected at the time---the new selection will automatically deselect any previously activated selector. This prevents an error caused by the system being unable to recognize about which axis the scatter plot should be rotated.

\item \textit{Help users recognize, diagnose, and recover from errors}: There are not many errors that user can commit, assuming the dataset is correct. The countermeasures against plotting incomplete data are built-in into the system. However, due to the nature of the visualization, erroneous blade geometries or any anomalies in their shapes will be detected by the user in a close-up inspection possible through the mixture of movement and maneuvering in the 3D space. The same can be said about the 3D scatter plots where user is able to rotate them and see them from every direction and can zoom in and zoom out from any data point using the same techniques.

\item \textit{Recognition rather than recall} and \textit{Aesthetic and minimalist design}: The blade's geometry is visualized as a replication of its physical appearance in the real-world. Moreover, both plots use volumetric glyphs to denote the points which reassembles the 2D scatter plots versions (see Fig.~\ref{fig:scatter_plot}). Previously selected data points are also highlighted. Furthermore, the system was designed to be minimalistic---only the efficiency and pressure ratio plots together with the geometry visualization are included to avoid overloading the user with information. Hence, for instance, the values of the data points are not initially visible, however, the scatter plots offer a possibility of gaining a high-level understanding of the data at a first glance. More detail of each individual point is available on an on-demand basis when the gaze cursor hovers over a data marker.

\item \textit{Help and documentation}: A succinct single page documentation sheet describing the interaction techniques is provided to the users.
\end{itemize}

\subsection{Expressiveness}
The expressiveness of the system is here analyzed using the cognitive dimensions of notation framework \cite{green_cognitive_1989}. This framework provides a vocabulary for analyzing the possibilities and limitations of an interactive notational system. Below we articulate how the keywords in this vocabulary maps onto the expressiveness of the system.

\begin{itemize}

\item \textit{Closeness of mapping}: The visualized geometry is a detail mapping of how would the blade would look like in the real-world.

\item \textit{Consistency}: All interactions are consistently designed and all interactive objects have consistent interaction qualities, such selectors.

\item \textit{Diffuseness / terseness}: The number of used symbols is minimized by only using  sphere-like markers with their characteristics (such as color, size and relative placement) to denote all the interactive elements of the visualization.

\item \textit{Error-proneness}: Errors are prevented using prevention mechanisms against error states, such  as selection of multiple rotation-selectors.

\item \textit{Hard mental operations}: Cognitive load and mental demand is kept to a minimum with straight-forward interaction methods and use of comparative visualizations.

\item \textit{Hidden dependencies}: As there is a 1:1:1 mapping between the visualization elements all the interdependencies are easily observed by the user since selection of a data point on one of the plots leads to a simultaneous selection of a corresponding data point on the other plot as well and the visualization of the associated geometry.

\item \textit{Juxtaposability} and \textit{Visibility}: All three main parts of the visualization, that is, the efficiency and pressure ratio plots together with the blade's geometry, are initially placed next to each other. Furthermore, the plots can be freely rearranged in the space as a group or individually. Moreover, the selection of any data point in a plot is automatically mapped on to the other plot as well and the corresponding geometry is immediately visible. In addition, selected data points are clearly visible through change in color and luminosity.

\item \textit{Premature commitment}: The user's workflow with the system is flexible and a user is free to initially inspect the nominal geometry, any or both plots, or to immediately select a data point.


\item \textit{Progressive evaluation}: Since all the user's actions result in immediate visual feedback (closed-loop interaction) the progress of the visual analytics task can be evaluated by the user at any time.

\item \textit{Role-expressiveness}: The individual roles of the three components, that is, the efficiency and pressure-ratio plots and the blade's geometry, are clear from the beginning, especially if the system is used by a domain expert.




\end{itemize}



\section{CONCLUSIONS AND FUTURE WORK} \label{sec:future}
The goal in this paper has been to introduce the \textit{AeroVR} system---a novel VR aerospace design environment with a particular emphasis on dimension reduction. We have identified the main structures of the design environment and implemented a fully working system for commodity VR headsets. We have also verified the interface from two perspectives: usability and expressiveness.

The two main identified tasks were (i) gaining the overview over the design and (ii) comparing the nominal geometry with the one associated with a specific performance parameters. The former was achieved through a mixture or visualization (e.g., blade and engine geometry, and scatter plots) and lower-level tasks (e.g., movement and interaction). The latter was achieved through visualization of the overlaying geometries: semi-transparent nominal blade superimposing over solid shape of a new design.  

Moving forward, our goal is to undertake the complete 3D design of a turbomachinery component in VR. In addition to the sufficient summary plots, our goal is to incorporate characteristics of blade performance at multiple operating points and have reduced order models to estimate the performance characteristics of new designs. 

In forthcoming years, the cost of the headset, and the required computing resources, will be further minimized with the introduction of next-generation of controllers, wireless headsets (e.g., Oculus Quest\cite{oculus}), and gestures (e.g., Leap Motion\cite{leapmotion}), or speech-based interfaces. Simultaneously, these rapid advancements in hardware and software open up completely new possibilities in terms of interaction techniques, rapid information analysis and the amount of data that can be processed and visualized at once. The two most promising venues of further development of the \textit{AeroVR} are investigation of which interaction techniques may bring the most benefit to the user and integration of our system with a system operating on knowledge from domain expert, i.e.~a~\textit{knowledge-based system}\cite{brodie_kbms_1989, nalepa_modeling_2018}. The former would include adding either controller-based laser-pointing or hand-tracking capabilities or a combination of thereof depending of the particular user's needs. The latter would require to develop and integrate a \textit{knowledge-(data)base}\cite{brodie_kbms_1989} with the interface provided by our system to build a \textit{knowledge-based system}\cite{brodie_kbms_1989, nalepa_modeling_2018}. 

\section*{ACKNOWLEDGEMENTS}
This work was supported by studentships from the Engineering and Physical Sciences Research Council (EPSRC-1788814), and the Cambridge European \& Trinity Hall Scholarship. The authors are grateful for all the generous support. The authors would also like to thank Timoleon Kipouros for his valuable suggestions.

\bibliographystyle{abbrv}
\bibliography{references}

\end{document}